\documentstyle[aps,prl,t1enc,epsf]{revtex}           %
\def\hs{\,-\,}
\begin{document}
\draft \twocolumn[\hsize\textwidth\columnwidth\hsize\csname@twocolumnfalse\endcsname
\preprint{}
%%%%%%%%%%%%%%%%%%%%%%%%%%%%%%%%%%%%%%%%%%%%%%%%%%%%%%
\title{Exact Inflation Braneworlds}
%%%%%%%%%%%%%%%%%%%%%%%%%%%%%%%%%%%%%%%%%%%%%%%%%%%%%%
\author{Deon Solomons\,, 
Peter Dunsby and George Ellis}
\address{Department of Mathematics and Applied Mathematics, University
  of Cape Town, \\ Rondebosch 7701, Cape Town, South Africa}
\maketitle
%%%%%%%%%%%%%%%%%%%%%%%%%%%%%%%%%%%%%%%%%%%%%%%%%%%%%%
\begin{abstract}
%%%%%%%%%%%%%%%%%%%%%%%%%%%%%%%%%%%%%%%%%%%%%%%%%%%%%%
In Randall\hs Sundrum type braneworld cosmologies, the dynamical 
equations on the three\hs brane differ from the general relativity 
equations by terms that carry the effects of embedding and of 
the free gravitational field in the five\hs dimensional bulk. 
In a FRW ansatze for the metric, we present two methods for 
deriving inflationary solutions to the covariant non\hs linear 
dynamical equations for the gravitational and matter fields 
on the brane. In the first approach we examine the constraints 
on the dynamical relationship between the cosmological scale factor 
and the scalar field driving self\hs interaction potential, imposed by 
the weak energy condition. We then investigate inflationary 
solutions obtained from a scalar field superpotential. Both these 
techniques for solving the braneworld field equations are illustrated 
by flat curvature models.
\end{abstract}
\vskip 1pc 
%\pacs{98.80.Cq \ \ 04.50.+h}
]
%%%%%%%%%%%%%%%%%%%%%%%%%%%%%%%%%%%%%%%%%%%%%%%%%%%%%%
\section{Introduction}\label{s:1}
%%%%%%%%%%%%%%%%%%%%%%%%%%%%%%%%%%%%%%%%%%%%%%%%%%%%%%
Motivated by developments in Superstring and M\hs theory 
\cite{witten,RS}, much literature has been devoted recently 
to studying the possibility that our observed universe 
may be viewed as a domain wall embedded in a higher dimensional 
space. It is assumed that in this scenario, the standard 
model interactions are confined to a $1+3$ dimensional 
hypersurface (referred to as the {\it the brane}), 
but the gravitational field may propagate through the {\it bulk} 
dimensions perpendicular to the brane. This view of cosmology has 
important consequences for the standard model of cosmology and in 
particular for the inflationary paradigm \hs which has successfully
explained many of the standard cosmological problems.
   
Braneworlds with a homogeneous and isotropic cosmology have recently
been explored by a number of authors \cite{FRW}
and in particular, solutions of the Friedmann\hs Robertson\hs Walker 
(FRW) dilatonic braneworlds have been investigated by Youm \cite{Youm}. 

Recent work on inflation \cite{MWBH} has demonstrated that the modified 
braneworld Friedmann equation leads to a stronger condition for inflation 
that amounts to violation of the strong energy condition. At high energies, 
feedback from the bulk dampens the rolling of the scalar field, in effect 
easing the condition for slow\hs rolling inflation for a given 
self\hs interaction, and increasing the number of e\hs foldings. 
Furthermore, the perturbation spectral index is driven towards 
the Harrison\hs Zel'dovich value $n=1$ \hs a very desirable feature 
of this model. 

However in the non\hs minimal coupling case it is not possible to derive 
an inflationary model from a string theory effective action on a 
generic background. This coupling slows down the expansion 
of the universe, thus spoiling the solution to the problems of 
large\hs scale structure formation and large\hs scale temperature anisotropy, 
both reasons for which inflation has been originally invoked \cite{nonmin}.

An algorithm for generating a class of exact braneworld inflationary models 
has been recently presented by Hawkins and Lidsey \cite{HL}, in the 
case where the {\it dark radiation} component has been 
neglected \footnote{This is motivated by the fact that the dark radiation 
rapidly redshifts away and soon becomes dynamically unimportant.}. 
It follows that the field equations have a simple 
{\it Hamilton\hs Jacobi} form and Hubble variable, which is written 
as a function of the scalar field, is then used as a solution\hs generating 
function for the self\hs interaction and scale factor evolution $a(t)$
(as proposed in \cite{fs,L}).   

Adopting the proposal by Randall and Sundrum \cite{RS}, where  
the dynamical equations on the three\hs brane differ from the general 
relativity equations by terms that carry the effects of embedding 
and of the free gravitational field in the five\hs dimensional bulk, 
we follow a two\hs tier approach \footnote{See \cite{HL,fs,L,me} 
where this program was first outlined in detail.} for finding 
exact inflationary FRW solutions to the Braneworld Einstein field 
equations that does not explicitly assume the slow\hs rolling approximation. 

We present two ways in which exact solutions to the covariant non\hs 
linear dynamical equations for the gravitational and matter fields on 
the brane can be obtained. 
In the first approach, which is based on a paper by Ellis and Madsen 
\cite{me}, we examine the constraints on the dynamical 
relationship between the cosmological scale factor and the scalar field 
driving self\hs interaction potential, imposed by the weak energy condition. 
The second approach then investigates inflationary solutions obtained 
from a scalar field superpotential (applied recently to the case of 
standard inflation by Feinstein\cite{fs}). Both these techniques of 
solving the braneworld field equations are illustrated by flat 
curvature models.
%%%%%%%%%%%%%%%%%%%%%%%%%%%%%%%%%%%%%%%%%%%%%%%%%%%%%%%%%
\section{Preliminaries}
%%%%%%%%%%%%%%%%%%%%%%%%%%%%%%%%%%%%%%%%%%%%%%%%%%%%%%%%%%
The 5\hs dimensional field equations\cite{RS,Maartens} 
are Einstein's equations, with a (negative) bulk cosmological 
constant $\widetilde{\Lambda}$ and the brane energy\hs 
momentum providing the source:
\begin{equation}
\widetilde{G}_{AB} =
\widetilde{\kappa}^2\left[- \widetilde{\Lambda}\widetilde{g}_{AB}
+\delta(\chi)\left\{
-\lambda g_{AB}+T_{AB}\right\}\right]\,. \label{1.1}
\end{equation}
The tildes denote the bulk (5\hs dimensional) generalization of
standard general relativity quantities, and
$\widetilde{\kappa}^2= 8\pi/\widetilde{M}_{\rm p}^3$, where
$\widetilde{M}_{\rm p}$ is the fundamental 5\hs dimensional Planck
mass, which is typically much less than the effective Planck mass
on the brane, $M_{\rm p}=1.2\times 10^{19}$ GeV. The brane is
given by $\chi=0$, so that a natural choice of coordinates is
$x^A=(x^\mu,\chi)$, where $x^\mu=(t,x^i)$ are spacetime
coordinates on the brane. The brane tension is $\lambda$, and
$g_{AB}=\widetilde{g}_{AB}-n_An_B$ is the induced metric on the
brane, with $n_A$ the spacelike unit normal to the brane. Matter
fields confined to the brane make up the brane energy\hs momentum
tensor $T_{AB}$ (with $T_{AB}n^B=0$).

The modification to the standard Einstein equations\cite{Maartens}, 
with the new terms carrying bulk effects onto the brane are 
\begin{equation}
G_{\mu\nu}=-\Lambda g_{\mu\nu}+\kappa^2
T_{\mu\nu}+\widetilde{\kappa}^4S_{\mu\nu} - {\cal E}_{\mu\nu}\,,
\label{1.2}
\end{equation}
where $\kappa^2=8\pi/M_{\rm p}^2$. We choose units in which 
$\kappa^2=1$, and $\lambda=6\left({\kappa\over\widetilde{\kappa}^2}\right)$. 
Here the tensor $S_{\mu\nu}$ represents local quadratic energy\hs momentum 
corrections of the matter fields, and the ${\cal E}_{\mu\nu}$ is the 
projected bulk Weyl tensor transmitting non\hs local gravitational degrees 
of freedom from the bulk to the brane. All the bulk corrections may be 
consolidated into effective total energy density $\rho^{tot}$, 
pressure $p^{tot}$, anisotropic stress $\pi^{tot}_{\mu\nu}$ 
and energy flux $q^{tot}_{\mu}$, as follows. The modified Einstein 
equations take the standard Einstein form with a redefined energy\hs 
momentum tensor:
\begin{equation}
G_{\mu\nu}=-\Lambda g_{\mu\nu}+\kappa^2 T^{\rm tot}_{\mu\nu}\,,
\label{25aa}
%\label{1.3}
\end{equation}
where
\begin{equation}
T^{\rm tot}_{\mu\nu}= T_{\mu\nu}+{\widetilde{\kappa}^{4}\over
\kappa^2}S_{\mu\nu}- {1\over\kappa^2}{\cal E}_{\mu\nu}\,.
\label{25bb}
%\label{1.4}
\end{equation}
Then
\begin{eqnarray}
\rho^{\rm tot} &=& \rho+{\widetilde{\kappa}^{4}\over
\kappa^6}\left[{\kappa^4\over24}\left(2\rho^2 -3
\pi_{\mu\nu}\pi^{\mu\nu}\right) +{\cal U}\right] \label{a}\\
%\label{1.5}\\
p^{\rm tot} &=& p+ {\widetilde{\kappa}^{4}\over
\kappa^6}\left[{\kappa^4\over24}\left(2\rho^2+4\rho p+
\pi_{\mu\nu}\pi^{\mu\nu}-4q_\mu q^\mu\right) +{\textstyle{1\over
3}}{\cal U}\right] \label{b}\\ 
%\label{1.6}\\
\pi^{\rm tot}_{\mu\nu} &=&\pi_{\mu\nu}+ {\widetilde{\kappa}^{4}\over
\kappa^6}\left[{\kappa^4\over12}\left\{-(\rho+3p)\pi_{\mu\nu}+
\pi_{\alpha\langle\mu}\pi_{\nu\rangle}{}^\alpha+ q_{\langle\mu}
q_{\nu\rangle}\right\}\right]\nonumber\\
& + & {\widetilde{\kappa}^{4}\over \kappa^6}{\cal P}_{\mu\nu}\label{c}\\
%\label{1.7}\\ 
q^{\rm tot}_\mu &=&q_\mu+ {\widetilde{\kappa}^{4}\over
\kappa^6}\left[{\kappa^4\over24}\left(4\rho
q_\mu-\pi_{\mu\nu}q^\nu\right)+ {\cal Q}_\mu\right] \,,\label{d}
%\label{1.8}\\
\end{eqnarray}
with non\hs local energy flux ${\cal Q}_\mu$ and non\hs local 
anisotropic stress ${\cal P}_{\mu\nu}$ both on the brane. 
(Note that $\widetilde{\kappa}^{4}/\kappa^6$ is dimensionless.)

These general expressions simplify in the case of a perfect fluid
(or minimally coupled scalar field, or isotropic one\hs particle 
distribution function), i.e., for
\begin{equation}
q_\mu=0=\pi_{\mu\nu}\;. 
\end{equation}
However, the total energy flux and anisotropic stress do not necessarily vanish: \[q^{\rm tot}_\mu ={\widetilde{\kappa}^{4}\over \kappa^6}{\cal
Q}_\mu\,,~~ \pi^{\rm tot}_{\mu\nu} ={\widetilde{\kappa}^{4}\over
\kappa^6}{\cal P}_{\mu\nu}\,. \]
%%%%%%%%%%%%%%%%%%%%%%%%%%%%%%%%%%%%%%%%%%%%%%%%%%%%%%
\section{The FRW Braneworld}\label{ss:Brane}\label{ss:2.2}
%%%%%%%%%%%%%%%%%%%%%%%%%%%%%%%%%%%%%%%%%%%%%%%%%%%%%%
\subsection{Matter description}
%%%%%%%%%%%%%%%%%%%%%%%%%%%%%%%%%%%%%%%%%%%%%%%%%%%%%%
A homogeneous scalar field $\phi(t)$ on the brane has an 
energy\hs momentum tensor  
\begin{equation}
T_{\mu\nu}=\rho u_\mu u_\nu+ph_{\mu\nu}\,,
 \label{22b}
\end{equation}
with $\rho={1\over 2}\dot{\phi}^2+V(\phi)$ and $p={1\over
  2}\dot{\phi}^2-V(\phi)$ where $\dot{\phi}^2$ is the kinetic energy
and $V(\phi)$ is the potential energy. The effective total 
energy\hs momentum tensor then also has
\begin{eqnarray}
\rho^{\rm tot} &=& \rho+{36\over
\lambda^2}\left[{1\over12}\rho^2 +{\cal U}\right] \label{22c}\\
p^{\rm tot} &=& p+ {36\over
\lambda^2}\left[{1\over12}\rho(\rho+2p) +{\textstyle{1\over
3}}{\cal U}\right]\;. \label{22d} 
\end{eqnarray}
A FRW braneworld is characterised by vanishing shear, acceleration, vorticity, 
non\hs local anisotropic stress and non\hs local energy flux 
(see section V equation (61) in \cite{Maartens}):
\begin{equation} 
\sigma_{\mu\nu}=A_{\mu}=\omega_{\mu}={\cal P}_{\mu\nu}={\cal Q}_\mu=0\;, 
\end{equation}
then
\begin{eqnarray}
\pi^{\rm tot}_{\mu\nu} &=&0\label{22e}\\ 
q^{\rm tot}_\mu &=&0 \,.\label{22f}
\end{eqnarray}
It follows that the standard conservation equation of general 
relativity holds, together with a simplified propagation equation 
for the non\hs local energy density ${\cal U}$ \cite{Maartens} 
\begin{eqnarray}
&&\dot{\rho}+\Theta(\rho+p)=0\,,\label{22g}\\ &&
\dot{\cal U}+{\textstyle{4\over3}}\Theta{\cal U}=0\,. \label{22h}
\end{eqnarray}
%%%%%%%%%%%%%%%%%%%%%%%%%%%%%%%%%%%%%%%%%%%%%%%%%%%%%%
\subsection{Gravitational  Equations}\label{ss:2.3}
%%%%%%%%%%%%%%%%%%%%%%%%%%%%%%%%%%%%%%%%%%%%%%%%%%%%%%
\newtheorem{define}{Definition}\newtheorem{theorem}{Theorem}
\newtheorem{proof}{Proof}\newtheorem{con}{Conjecture}
The Einstein Field equations with $\Lambda=0$ then become 
\begin{eqnarray}
&&{}\nonumber\\&&
\dot{\Theta}+ {\textstyle{1\over 3}}\Theta^2=-{\textstyle{1\over 2}}
(\rho^{\rm tot}+3p^{\rm tot}) \,,\label{23g}
\\&&{}\nonumber\\ &&
%3\dot{{\cal K}}+2\Theta{\cal K}=0\,,\label{23h}
%\\&&{}\nonumber\\ &&
{\textstyle{1\over 3}}\Theta^2+3{\cal K}=\rho^{\rm tot}\,,\label{23i}
\\&&{}\nonumber\\ && 
{36\over\lambda^2} {\cal U}=-\Gamma^2 a^{-4}\,,\label{23k}
\end{eqnarray}
where ${\cal K}=k/a(t)^2$, $k=0$, $\pm 1$ and $\Gamma$ is a 
arbitrary real constant.
%%%%%%%%%%%%%%%%%%%%%%%%%%%%%%%%%%%%%%%%%%%%%%%%%%%%%%
\section{Generating Exact Inflation Braneworlds \label{s:3}}
%%%%%%%%%%%%%%%%%%%%%%%%%%%%%%%%%%%%%%%%%%%%%%%%%%%%%%
\subsection{The Reality Condition}\label{ss:2.4}
%%%%%%%%%%%%%%%%%%%%%%%%%%%%%%%%%%%%%%%%%%%%%%%%%%%%%%
The effects of embedding the free gravitational field in the 
five\hs dimensional bulk are represented by the correction terms to the 
energy\hs momentum tensor $T_{\mu\nu}$, as expressed in equations (\ref{22c}) 
and (\ref{22d}). In the absence of anisotropic stress ($\pi_{\mu\nu}^{\rm tot}=0$) the 
non\hs local energy density ${\cal U}$ takes the form of {\it dark radiation} 
with its behaviour given by equation (\ref{23k}). 

As a first step to specifying a desired model, we note that the
following combination of field equations (\ref{23g})-(\ref{23i}) gives the 
sum of the effective (total) energy density and pressure:
\begin{equation}
\rho^{\rm tot}+p^{\rm tot}=2\left({\cal K}-
{\textstyle{1\over 3 }}\dot{\Theta}\right)
 \, .\label{24a}\end{equation} 
Adding equations (\ref{22c}) and (\ref{22d}) and factorizing the
resulting expression, we find
\begin{equation}  
\rho^{\rm tot}+p^{\rm tot}= \dot{\phi}^2\left(1+{\textstyle{6\over \lambda^2}}
\rho\right)
+{\textstyle{48\over \lambda^2}}
{\cal U}\;,
\end{equation}
and substituting this relation into (\ref{24a}) and rearranging terms 
we obtain an expression for $\dot{\phi}^2$:
\begin{equation}
\dot{\phi}^2\left(1+{\textstyle{6\over \lambda^2}}
\rho\right)=2\left({\cal K}-{\textstyle{1\over 3}}
\dot{\Theta}\right)
-{\textstyle{48\over \lambda^2}}
{\cal U} \, .\label{24b}\end{equation} 
\noindent {\bf Definition:} The Reality Condition holds iff $\dot{\phi}^2
\geq 0$. This statement is equivalent to the {\it Null Energy Condition} 
$\rho+p\geq 0$ and requires
\begin{equation}
\left(1+{\textstyle{6\over \lambda^2}} \rho\right)>0\label{RC.1}\\
\end{equation}
and
\begin{equation}
2\left({\cal K}-{\textstyle{1\over 3}} \dot{\Theta}\right)-
{\textstyle{48\over \lambda^2}}{\cal U} 
\geq  0\label{24e}
\end{equation}
to hold at all times $t$. The factor $\left(1+{\textstyle{6\over 
\lambda^2}} \rho\right)$ is determined by equations 
(\ref{22c}) and (\ref{23i}):
\begin{equation}
{\textstyle{3\over\lambda^2}}\rho^2+\rho+{\textstyle{36\over\lambda^2}}U
-{\textstyle{1\over3}}\Theta^2-3K=0\;.\label{26}
\end{equation}
In order to satisfy condition (\ref{RC.1}) we take the positive 
root of equation (\ref{26}):
\begin{equation}
1+{\textstyle{6\over \lambda^2}}
\rho
=\sqrt{1+{\textstyle{12\over \lambda^2}}
\left[(3{\cal K}+{\textstyle{1\over 3}}\Theta^2)
-{\textstyle{36\over \lambda^2}}{\cal U}\right]} \, .
\label{24c}\end{equation} 
Note that the Weak Energy Condition, viz. $\rho\geq 0$ 
{\em and} $\rho+p\geq 0$ holds if  
\begin{equation}3{\cal K}+{\textstyle{1\over 3}}\Theta^2
-{\textstyle{36\over \lambda^2}}{\cal U}\geq 0\label{24d}\end{equation} 
since this inequality ensures that $\rho$ is non\hs negative 
in equation (\ref{24c}). The Weak Energy Condition is 
violated if $\rho<0$, i.e.: 
\begin{enumerate}
\item if condition (\ref{24d}) does not hold,  
\item or if both inequalities (\ref{RC.1}) and (\ref{24e}) are 
reversed\footnote{Notwithstanding violation of the 
Weak Energy Condition, the Reality Condition may in fact 
still hold, due to this double change in sign.}. 
\end{enumerate}
%%%%%%%%%%%%%%%%%%%%%%%%%%%%%%%%%%%%%%%%%%%%%%%%%%%%%%
\subsection{How to obtain Braneworlds with desired inflationary behaviour}
\label{ss:3.1}
%%%%%%%%%%%%%%%%%%%%%%%%%%%%%%%%%%%%%%%%%%%%%%%%%%%%%%
We now present inflationary braneworld solutions based upon the 
Reality Conditions (\ref{RC.1}) {\em and} (\ref{24e}). Using equations (\ref{24b}) and (\ref{24c}) $\dot{\phi}^2$ can be written in the following useful form:
\begin{equation}
\dot{\phi}^2={2\over3}{\Phi\over\Delta}\;,
\label{a1}
\end{equation}
where
\begin{equation}
\Phi=-{\textstyle{48\over\lambda^2}}U+2({\cal K}-
{\textstyle{1\over3}}\dot{\Theta})
\label{a2}
\end{equation}
and
\begin{equation}
\Delta=
\sqrt{1+{\textstyle{12\over \lambda^2}}
\left({\textstyle{1\over 3}}\Theta^2+3{\cal K}
-{\textstyle{36\over \lambda^2}}{\cal U}\right)}\;.
\label{a3}
\end{equation}
A quadratic  equation for the  potential $V(\phi)$ is determined by taking 
the combination 2(\ref{23i})+(\ref{23g}) of the field equations and 
using equations (\ref{22c}) and (\ref{22d}):
\begin{equation}
V(\phi)^2+{\textstyle{\lambda^2\over3}}V(\phi)-{\textstyle{\lambda^2\over9}}
\left[\dot{\Theta}+\Theta^2+6{\cal K}+{\textstyle{1\over\lambda^2}}{\textstyle{\Phi^2\over\Delta^2}}-{\textstyle{36\over\lambda^2}}\right]=0\;.
\label{a4}
\end{equation}
Taking the positive root\footnote{This ensures consistency with 
general relativity in the limit $\lambda\rightarrow\infty$.} gives
\begin{equation}
V(\phi)={\lambda^2\over
  6}\left[\sqrt{1+{\textstyle{4\over\lambda^2}}\left(\dot{\Theta}+
\Theta^2+6{\cal K}+{\textstyle{1\over\lambda^2}}{\textstyle{\Phi^2\over\Delta^2}}-
{\textstyle{36\over\lambda^2}}{\cal U}\right)}-1\right]\;.
\label{a5}
\end{equation}
In order to obtain the desired inflationary dynamics we proceed as
follows. (i) Choose $k$ (and therefore determine ${\cal K}$); (ii) 
specify a (monotonic) function $a(t)$ as desired, provided 
the {\it Brainworld Reality Condition} is satisfied; (iii) determine
the expansion parameter $\Theta=3\dot{a}(t)/a(t)$ and its derivative 
$\dot{\Theta}$. Integrating equation (\ref{a1}) gives $\phi(t)$ and
inverting gives $t(\phi)$. Finally equation (\ref{a5}) gives
$V(t)=V(t(\phi))\Rightarrow V(\phi)$. Thus the above procedure will
determine $V(\phi)$ describing a model which satisfies the exact
inflationary equations on the brane (with no slow\hs rolling
approximation) with the desired scale factor $a(t)$ behaviour.

It should be pointed out that this procedure is consistent as all the 
above dynamical equations are satisfied at all times.
%%%%%%%%%%%%%%%%%%%%%%%%%%%%%%%%%%%%%%%%%%%%%%%%%%%%%
\subsection{Simple examples}
%%%%%%%%%%%%%%%%%%%%%%%%%%%%%%%%%%%%%%%%%%%%%%%%%%%%%%
\subsubsection{The de\hs Sitter model}\label{sss:3.11}
%%%%%%%%%%%%%%%%%%%%%%%%%%%%%%%%%%%%%%%%%%%%%%%%%%%%%%
There is a simple solution of equations (\ref{a1}\hs \ref{a5}) for
$U=0$ which yields the simple exponential inflationary solution 
representing a de\hs Sitter universe:
\begin{equation} 
a(t)\equiv \exp(\alpha(t))=\exp(\omega t)\;,~~\dot{\omega}=0\;,
\end{equation} 
corresponding  to a constant potential and stationary scalar field 
with no non\hs local energy density:
\begin{equation}
V(\phi)={\lambda^2\over 6}\left[\sqrt{1+{36\omega^2\over \lambda^2}}-1
\right]\,, \, \, \dot{\phi}(t)=0\,, \, \, {\cal U}=0\;.
\end{equation}
This solution asymptotes to the GR result \cite{me} 
as $\lambda\rightarrow\infty$, since $V\rightarrow 3\omega^2$.
An additional {\it dark radiation} de\hs Sitter inflationary solution
exists. However, it cannot be inverted so we chose to omit it here.
%%%%%%%%%%%%%%%%%%%%%%%%%%%%%%%%%%%%%%%%%%%%%%%%%%%%%%
\subsubsection{de\hs Sitter expansion from a singularity}\label{sss:3.12}
%%%%%%%%%%%%%%%%%%%%%%%%%%%%%%%%%%%%%%%%%%%%%%%%%%%%%%
Consider now inflationary de\hs Sitter expansion from a 
singularity ($a(0)=0$):
\begin{equation}
a(t)\equiv \exp(\alpha(t))=\sinh(\omega t)\;,~~\dot{\omega}=0\;,
\end{equation}
An exact solution exists if we choose the constant 
\begin{equation}
\Gamma^2=\textstyle{{27w^4\over p^2}}\, ,\ \ p^2=\lambda^2+36\omega^2\ .
\end{equation}
Now $\dot{\phi}^2={2\lambda \omega^2\over p}\ e^{-2\alpha}$, so that 
$\dot{\phi}=\pm \sqrt{2\lambda\over p}\ {\omega\over\sinh{(wt)}}$ .\\
Then we find (for $\phi(\infty)=\phi_{\infty}$)
\begin{eqnarray}
\phi &=& \phi_{\infty}\pm \sqrt{{2\lambda\over p}}\ln{\left|
\tanh{({\omega t\over 2})}\right|}\,, \nonumber\\
V(\phi) &=& 
{\lambda^2\over 6}\left(\sqrt{1+{36\omega^2\over \lambda^2}}-1\right)\\
&+& {2\omega^2\over \sqrt{1+{36\omega^2\over \lambda^2}}}\ 
\sinh^2{\left[\sqrt[4]{{1\over 4}\left(1+{36\omega^2\over \lambda^2}\right)}
(\phi-\phi_{\infty})\right]} .\nonumber
\end{eqnarray}
%%%%%%%%%%%%%%%%%%%%%%%%%%%%%%%%%%%%%%%%%%%%%%%%%%%%%%
\subsubsection{de\hs Sitter expansion from a bounce}\label{sss:3.13}
%%%%%%%%%%%%%%%%%%%%%%%%%%%%%%%%%%%%%%%%%%%%%%%%%%%%%%
We next consider inflationary de\hs Sitter expansion from a bounce:
\begin{equation}
a(t)\equiv\exp(\alpha(t))=\cosh(\omega t)\;,~~\dot{\omega}=0\;,  
\end{equation}
An exact solution exists if we choose 
\begin{equation}\Gamma^2=\textstyle{{27w^4\over p^2}}\; , 
\ p^2=\lambda^2+36\omega^2\;.\end{equation} 
Again $\dot{\phi}^2={2\lambda \omega^2\over p}\ e^{-2\alpha}$, so that 
$\dot{\phi}=\pm \sqrt{2\lambda\over p}\ {\omega\over\cosh{(wt)}}$. \\The 
scalar field solution (for $\phi(0)=\phi_0$) and the potential is
\begin{eqnarray}
\phi &=& \phi_0\pm\sqrt{{2\lambda\over p}}\tan^{-1}{(\sinh{\omega t})}
\,, \nonumber\\
V(\phi) &=& 
{\lambda^2\over 6}\left(\sqrt{1+{36\omega^2\over \lambda^2}}-1\right)\\
& - & {2\omega^2\over \sqrt{1+{36\omega^2\over \lambda^2}}}\ 
\cos^2{\left[\sqrt[4]{{1\over 4}\left(1+{36\omega^2\over 
\lambda^2}\right)}(\phi-\phi_0)\right]}\; .\nonumber  
\end{eqnarray}
%%%%%%%%%%%%%%%%%%%%%%%%%%%%%%%%%%%%%%%%%%%%%%%%%%%%%%%%%%
\subsubsection{The coasting solution}\label{sss:3.14}
%%%%%%%%%%%%%%%%%%%%%%%%%%%%%%%%%%%%%%%%%%%%%%%%%%%%%%%%%%
Finally let us consider linear expansion 
\begin{equation}
a(t)\equiv\exp(\alpha(t))=\omega\ t\;,~~\dot{\omega}=0\;. 
\end{equation}
An exact solution exists if we choose
\begin{equation}
\Gamma^2={27\omega^2\over \lambda^2}\;. 
\end{equation}
Then
\begin{eqnarray}
\phi(t)&=&\sqrt{2}\ln{|t|}\nonumber\,,\\
V(\phi)&=&2e^{-\sqrt{2}\phi}\;.
\end{eqnarray}
An interesting feature of this solution is that it is identical to the 
GR case (see equation (54) in \cite{me}).
%%%%%%%%%%%%%%%%%%%%%%%%%%%%%%%%%%%%%%%%%%%%%%%%%%%%%%%%%%
\section{Flat Braneworlds with desired scalar field self\hs interaction}
\label{ss:3.2}
%%%%%%%%%%%%%%%%%%%%%%%%%%%%%%%%%%%%%%%%%%%%%%%%%%%%%%%%%%
We now demonstrate how the superpotential formalism developed by 
Feinstein\cite{fs} and introduced in a more subtle way via the 
Hamilton\hs Jacobi equations in \cite{HL} also yield solutions 
to the braneworld field equations starting from the general anzatse 
$\Theta=\Theta(\phi)$. The first step is to ensure that the 
Reality Conditions (\ref{RC.1}) {\em and} (\ref{24e}) are met for 
any value of ${\cal K}$. We then derive a system of non\hs linear 
differential equations variable in $\phi$, that are equivalent 
to the Einstein field equations (\ref{23g})\hs (\ref{23k}). This 
system is solved for the flat curvature case ${\cal K}=0$, with 
the specific ansatze $\Theta=\theta_0e^{m\phi}$ 
($\dot{\theta_0}=0=\dot{m}$). In this way we recover the asymptotic 
power\hs law inflationary solution in the limit $\lambda\rightarrow\infty$, 
or alternatively, for $m\phi<<0$.
%%%%%%%%%%%%%%%%%%%%%%%%%%%%%%%%%%%%%%%%%%%%%%%%%%%%%%%%%%
\subsection{The Braneworld Superpotential}\label{ss:2.5}
%%%%%%%%%%%%%%%%%%%%%%%%%%%%%%%%%%%%%%%%%%%%%%%%%%%%%%%%%%
The superpotential ansatze is
$$\Theta\equiv\Theta(\phi)$$ in terms of the scalar field $\phi$. 
We start again with $\Delta$ defined by equation (\ref{a3}):
\begin{equation}
\Delta^2=1+{\textstyle{12\over \lambda^2}}
\left({\textstyle{1\over 3}}
\Theta^2+3{\cal K}-\textstyle{{36\over \lambda^2}}{\cal U}\right)\,,
\label{25a}\end{equation} 
then the local energy density is give by 
\begin{equation}
\rho=\textstyle{{\lambda^2\over 6}}(\Delta-1)\,,
\label{25b}
\end{equation} 
and therefore the kinetic energy of the scalar field  
$\psi=\dot{\phi}$ is a solution of the quadratic equation \footnote{ 
prime $()'$ denotes differentiation with respect to $\phi$.}
\begin{equation}\Delta\psi^2+{\textstyle{2\over 3}}\Theta'\psi+
{\textstyle{48\over\lambda^2}}{\cal U}-2{\cal K}=0
\,,\label{25c}\end{equation} and upon setting ${\cal K}=0$, and eliminating 
the dark radiation term ${\cal U}$ using equation (\ref{23k}), 
\begin{equation}\left(1+\textstyle{{\lambda^2\over 6}}
\rho\right)\psi^2+{\textstyle{2\over 3}}\Theta'\psi+
{\textstyle{4\over3}}
\left[\Theta^2-\rho\left(1+\textstyle{{\lambda^2\over 12}}\rho\right)
\right]=0
\,,\label{25d}\end{equation}
which yields two solutions for $\psi$. Together with energy 
conservation (\ref{22g}) 
\begin{equation}\rho'+\Theta\ \psi=0\,,\label{25d1}\end{equation}
one can obtain solutions for $\rho(\phi)$ and then for $\phi(t)$.
The solutions to equations (\ref{25d}) and (\ref{25d1}) are real for 
${\cal K}\geq 0$ (note that we assume ${\cal U}\leq 0$). There are 
{\em two} Reality Conditions that need to be satisfied in the case 
${\cal K}<0$, viz.
\begin{eqnarray}&&\textstyle{{1\over 9}}\Theta'^2+\Delta\left(2{\cal K}
-{\textstyle{48\over\lambda^2}}{\cal U}\right)\geq 0\, ,\label{Real1}
\\&&\nonumber\\
&&{\textstyle{1\over 3}}\Theta^2+3{\cal K}-
\textstyle{{36\over \lambda^2}}{\cal U}\geq 0\,.
\label{Real2}\end{eqnarray}
the first of which corresponds to the Reality Condition, 
while the second ensures that the RHS of equation (\ref{25b}), and therefore $\rho$, is positive. The non\hs linear differential 
equations for the non\hs local energy density and the 
scale factor $e^{\alpha(\phi)}$ are
\begin{eqnarray}
%&&{\cal K}'(\phi)=-{\textstyle{2\over3}}{\Theta\over 
%\psi_{\pm}}{\cal K}(\phi)\,,\label{25e}\\&&\nonumber\\
&&{\cal U}'(\phi)=-{\textstyle{4\over3}}
{\Theta\over \psi_{\pm}}{\cal U}(\phi)\,,\label{25f}\\&&\nonumber\\
&&\alpha'(\phi)=-{\textstyle{1\over3}}
{\Theta\over \psi_{\pm}}\,,\label{25g}
\end{eqnarray} with the temporal equation 
\begin{equation}t':={1\over \psi_{\pm}}\,.\label{25h}
\end{equation} 
%%%%%%%%%%%%%%%%%%%%%%%%%%%%%%%%%%%%%%%%%%%%%%%%%%%%%%%%%%
\subsection{Power\hs law inflation}\label{sss:3.21}
%%%%%%%%%%%%%%%%%%%%%%%%%%%%%%%%%%%%%%%%%%%%%%%%%%%%%%%%%%
For flat curvature the Reality Condition (\ref{Real1}) and condition
(\ref{Real2}) are trivially satisfied. 
%We demonstate how the superpotential 
%formalism developed by Feinstein\cite{fs} {\em without} `dark radiation' 
%yields solutions to the braneworld field equations starting from the 
%anzatse 
%$\Theta=\theta_0e^{m\phi}$ for constants $\theta_0$ and $m$. 
%We then recover the asymptotic power\hs law inflationary solution 
%at late times. 
In the absence of {\it dark radiation}, equation (\ref{25d}) reduces to
\begin{equation}
\left(1+{6\over\lambda^2}\rho\right)\psi^2+{\textstyle {2\over
    3}}\Theta'\psi=0\;.\label{sss321aa}
\end{equation}
Now if $\psi=0$, we recover the {\em de Sitter} solutions of section (\ref{sss:3.11}) provided that $\rho=constant>0$. For $\psi\neq 0$, equation ({\ref{sss321aa}) together with (\ref{25d1}), leads to the system
\begin{eqnarray}
\dot{\phi}&=&{-2\Theta'\over 
3\left(1+{6\over\lambda^2}\rho\right)}\, ,\label{sss321a}\\
\rho'&=&-\Theta\psi \, .\nonumber\end{eqnarray}
If we assume that 
\begin{equation}
\Theta=\theta_0e^{m\phi} 
\end{equation}
for constants $\theta_0$ and $m$, the local energy density 
\begin{equation}
\rho={\lambda^2\over 6}\left(\sqrt{1+{4\theta_0^{\ 2}\over 
\lambda^2}e^{2m\phi}}-1\right)\,,
\end{equation}
and (\ref{sss321a}) integrates to give
\begin{eqnarray}
{2m^2\theta_0\over 3}t&=&\sqrt{{4\theta_0^{\ 2}\over \lambda^2}+e^{-2m\phi}}
-{2\theta_0\over \lambda}m\phi\nonumber\\
&-&{2\theta_0\over \lambda}
\ln{\left[{2\theta_0\over \lambda}+\sqrt{{4\theta_0^{\ 2}\over \lambda^2}
+e^{-2m\phi}}\right]}\;.
\end{eqnarray}
This asymptotes to $e^{-m\phi}$ for $m\phi\ll 0$. Alternatively, we recover 
the General Relativity  result (see \cite{me}) in the limit 
$\lambda\rightarrow \infty$. This asymptotic solution implies 
that $\Theta\sim {3\over 2m^2\ t},$ which in turn means that 
\begin{equation}
a(t) \sim t^{1/2m^2}
\end{equation}
i.e., the model exhibits power\hs law inflation 
for $m^2<{1\over 2}$. 

The potential is then obtained using 
\begin{equation}
V(\phi)=\rho-{\textstyle 1\over 2}\dot{\phi}^2 
\end{equation}
together with equation (\ref{sss321a}) and has the form
\begin{eqnarray}
& V(\phi) = \textstyle{{\lambda^2\over 6}}\left[\sqrt{1+
\textstyle{{4\theta_0^{\ 2}\over \lambda^2}}e^{2m\phi}}-1\right] 
&\nonumber\\ & -\textstyle{{2\over 9}}m^2\theta_0^{\ 2}e^{2m\phi}\left(1+
\textstyle{{4\theta_0^{\ 2}\over \lambda^2}}e^{2m\phi}\right)^{-1}&
\end{eqnarray}
which asymptotes to 
$V\sim \textstyle{{3-2m^2\over 9}}\theta_0^{\ 2}e^{2m\phi}$ during 
power\hs law inflation.
%%%%%%%%%%%%%%%%%%%%%%%%%%%%%%%%%%%%%%%%%%%%%%%%%%%%%%%%%%
\section{Conclusion}
%%%%%%%%%%%%%%%%%%%%%%%%%%%%%%%%%%%%%%%%%%%%%%%%%%%%%%%%%%
In conclusion, we have presented two algorithms for solving
the braneworld generalisations of the FRW equations for a single
classical non\hs minimally coupled scalar field which is confined 
to the brane, embedded in five dimensional Einstein gravity. 
The first approach, which mirrors the one followed by Madsen 
and Ellis \cite{me}, does not assume any slow rolling approximation 
however breaks down when the field oscillates about the minimum 
of the potential well. The second approach extends recent work 
by Feinstein \cite{fs} where a superpotential method, inspired 
by studies in supergravity, is used to generate standard inflationary 
solutions.  

The utility of these algorithms is demonstrated through a number of 
simple examples including the Braneworld analogues of de\hs Sitter 
exponential inflation and power\hs law inflation. All our solutions
reduce to the standard result in the appropriate limit.

%%%%%%%%%%%%%%%%%%%%%%%%%%%%%%%%%%%%%%%%%%%%%%%%%%%%%%%%%%

\end{document}